\documentclass[
aps,
prl,
reprint,
amsfonts,
amssymb,
amsmath
]{revtex4-2}
\usepackage{natbib}
\usepackage{bm,latexsym,mathrsfs,enumitem}
\usepackage{upgreek}
\usepackage[table,x11names]{xcolor}
\usepackage[breaklinks=true,
unicode=true,
urlcolor = RoyalBlue4,
colorlinks = true,
citecolor = RoyalBlue4,
linkcolor = RoyalBlue4
]{hyperref}
\usepackage{graphicx}
\graphicspath{{./figs/}}
\usepackage{mathtools}
\usepackage[cal=boondoxo,frak=boondox,scr=rsfs]{mathalfa}
\mathtoolsset{showonlyrefs=true}
\usepackage{savesym}
\usepackage[most]{tcolorbox}
\savesymbol{comment}
\usepackage[todonotes={textsize=tiny,textwidth=47pt},truncate=hyphenate,deletedmarkup=xout,commandnameprefix=ifneeded]{changes}
\definechangesauthor[name={Denis Sheka},color=DodgerBlue1]{DS}
\definechangesauthor[name={Olek},color=red]{Olek}
\definechangesauthor[name={Gaetano Napoli},color=blue]{GN}
\definechangesauthor[name={Luigi Vergori},color=magenta]{LV}
\usepackage{gitinfo2}
\def\mv{{\bf m}}
\def\nv{{\bf n}}
\def\Nv{{\bf N}}

\newcommand{\ba}{\begin{eqnarray}}
\newcommand{\ea}{\end{eqnarray}}
\newcommand{\be}{\begin{equation}}
\newcommand{\en}{\end{equation}}
\def\Nv{{\bf N}}

\def\ev{{\bf e}}

\def\nv{{\bf n}}

\def\mv{{\bf m}}

\def\Mv{{\bf M}}

\def\Iv{{\bf I}}
\def\Qv{{\bf Q}}

\def\eps{ \varepsilon}

\def\grads{\nabla\hspace{-1mm}_s}

\def\esse{{{S}}}

\newcommand{\ot}{\otimes}

\def\bnu{\boldsymbol{\nu}}

\def\w0{{\mathscr{E}_0}}
\def\wa{{\mathscr{E}_{\textsc{A}}}}
\def\wdmi{{\mathscr{E}_{\textsc{DM}}}}

\begin{document}
\preprint{\textcolor[rgb]{0.00,0.50,0.75}{{\texttt{Version \gitAbbrevHash{} by \gitCommitterName{} on \gitCommitterIsoDate}}}}	

\title{Nematic versus ferromagnetic shells: new insights in curvature-induced effects}

\author{Gaetano Napoli}
\email{gaetano.napoli@unisalento.it}
\affiliation{Dipartimento di Matematica e Fisica "E. De Giorgi", Universit\`a del Salento, Lecce (Italy).}
\author{Oleksandr V. Pylypovskyi}
\email{o.pylypovskyi@hzdr.de}
\affiliation{Helmholtz-Zentrum Dresden-Rossendorf e.V., Institute of Ion Beam Physics and Materials Research, 01328 Dresden, Germany}
\affiliation{Kyiv Academic University, 03142 Kyiv, Ukraine}
\author{Denis~D.~Sheka}
\email{sheka@knu.ua}
\affiliation{Taras Shevchenko National University of Kyiv, 01601 Kyiv, Ukraine}

\author{Luigi Vergori}
\email{luigi.vergori@unipg.it}
\affiliation{Dipartimento di Ingegneria, Universit\`a di Perugia, Perugia (Italy)
}%

\date{\today}

\begin{abstract}
 Within the framework of continuum theory, we draw a parallel between ferromagnetic materials and nematic liquid crystals confined on curved surfaces,  which are both characterized by local interaction and anchoring potentials.  {We show that the extrinsic curvature of the shell combined with the out-of-plane component of the director field gives rise to chirality effects. This interplay produces an effective energy term reminiscent of the chiral term in cholesteric liquid crystals, with the curvature tensor acting as a sort of anisotropic helicity.   {We discuss also} how the different nature of the order parameter, a vector in ferromagnets and  a tensor in nematics,  yields different textures on  surfaces with the same topology as the sphere.}  In particular, 	 {we show that the extrinsic curvature governs the ground state configuration on a nematic spherical shell,  favouring two antipodal disclinations of charge +1 on small particles and four $+1/2$ disclinations of charge  located at the vertices of a square inscribed in a great circle on larger particles.}
  
\end{abstract}

\maketitle

 {Magnetic materials and liquid crystals are examples of materials with orientational order which give rise to textures whose complexity is as beautiful as challenging to study. Their confinement to curved layers causes the emergence of geometry-induced effects that are not usually observed in flat layers.}
Recent  advances in the theory of curvilinear magnetism have highlighted a range of fascinating geometry-induced effects in the magnetic properties of materials \cite{Streubel16a,Vedmedenko20}. When confined to thin curved domains, effective physical features arise from  the interplay between the curved geometry and magnetic texture. 

According to continuum micromagnetic description of the ferromagnetic media, the magnetic textures can be well described by the vector order parameter $\mv =\Mv/ |\Mv|$, which is the normalized magnetization vector. The energy of the ferromagnet typically includes:  (i) a short-range exchange energy that penalizes the non-uniformity of the magnetization; (ii) an anisotropic term that models the existence of directions preferred by the magnetization; (iii) an non-local term describing the long-range magnetostatic interactions. 

When ferromagnets are confined in thin curvilinear layers, the magnetic energy can be decomposed \cite{Sheka:2020} to reveal the emergence of geometry-induced anisotropy and geometry-induced chiral interaction with emergent Dzyaloshinskii--Moriya interaction (DMI) as characteristic example.  With this decomposition,  a number of new effects in ferromagnetic spherical shells have been studied  including topological patterning and magnetochiral effects, for the review see Refs.~\cite{Streubel16a,Fischer20,Vedmedenko20}. Moreover,
recent advances in experimental techniques have also made possible  the manipulation of these effects for the design of new functional materials and applications for spintronics, shapeable magnetoelectronics, magnonics, biomedicine, and soft robotics \cite{Makarov16,Streubel16a,Fischer20,Vedmedenko20} .

Soft matter also provides  an area in which the interplay between geometry of the substrate and the order parameter  plays a crucial role. One example  is provided by liquid crystal (LC)  shells \cite{Lopez-Leon11}.  These are microscopic colloidal  particles coated with a thin layer of nematic LC, and  have potential applications as the topological defects (which may occur on them) can be engineered to emulate the linear, trigonal and tetrahedral geometries of carbon atoms \cite{Nelson:2002}. This feature opens up the possibility to design meso-atoms  with special optical properties whose valence and directional-binding can be controlled. Photonic lattices made of LC shells are the new frontier for the manufacture of a new-generation optical cryptographic devices \cite{Geng:2016}.

Nematic LCs are aggregates of rodlike molecules. Within the classic theory  of nematics\cite{virga}, the average microscopic molecular orientation is described by a sole vector order parameter $\nv$ called  the \emph{director}.  In the 1960s  de Gennes introduced  the order-tensor theory which bases on the orientational probability distribution and provides measures of the degree of orientation and biaxiality. In its simplest form, this theory uses as state variable a second-order symmetric traceless tensor $\Qv = s(\nv \ot \nv - \frac{1}{3} \Iv)$, with  the scalar parameter $s$ being the {\it degree of orientation} that  vanishes at points where there is no privileged direction. Contrarily to the director theory, de Gennes theory allows the study of nematic-isotropic phase transitions.

Most theories of nematic shells are based on energy functionals defined on surfaces expressed in terms of vector \cite{Vitelli:2006, Napolia:2012} or tensor \cite{Biscari:2009, Kralj:2011, Napolib2012} order parameters. Theories in which the director field $\nv$ is purely tangential have the flaw that  topological defects, i.e. points where the director is not uniquely defined, inevitably arise on surfaces with the same topology as sphere.  Unavoidably, the energy blows up in a neighbourhood  of a defect.  This flaw can be overcome by introducing a theory  in which the tensor order parameter $\Qv$ is tangential, i.e. the normal to the surface is an eigenvector of $\Qv$ corresponding to the zero eigenvalue. In this framework, {\it melting points}  (points at which the degree of orientation vanishes) occur, and  the energy is finite over the entire domain. A different way to avoid singularities within the director theory is to release the constraint that the director field is tangential (as suggested by recent experimental observations \cite{Durey:2020, Noh:2020}), and to  assign an energy cost to the out-of-plane components of $\nv$.  {This hypothesis,  first proposed  in  \cite{Cladis:1972}, allows to deal with  smooth fields as topological defects revert into non-singular disclinations.} 
This alternative will be explored in this paper and will allow us to compare the theories of ferromagnetic and nematic shells.  {Note that in a more comprehensive framework, fusion of the defect core and out-of-plane escape can be considered simultaneously \cite{Susser:2020}, as well as biaxial order reconstruction \cite{Schopohl:1987, Carbone:2009}.}

In theories based on vector order parameters in which a privileged  direction is defined everywhere on the shell,  {due to the different {\it orientability} of the state variables, there is a fundamental difference between the response of ferromagnets and nematics.} For nematics, the head-tail symmetry of the molecules is translated at the macroscopic level by the assumption that $\nv$ must be indistinguishable from $-\nv$.
Roughly speaking, it is more appropriate to regard the director as a {\it line}  instead of an {\it arrow} \cite{Ball:2008, Ball:2011}. We may then conclude that no vector order parameter can describe properly the mechanical response of nematics. A tensor order parameter has to be introduced. For this reason, in our attempt to  {determining} the equilibrium textures of a nematics on a curved shell $\esse$, we  {choose to} use the second-order alignment tensor $\Nv = \nv \ot \nv$ instead of the director field $\nv$. Introduced the local orthonormal basis $\mathcal{B}=\{\ev_1,\ev_2,\bnu\}$, where $\ev_1$ and $\ev_2$ are the principal directions on  $\esse$ and $\bnu=\ev_1\times\ev_2$ is the unit normal to $\esse$, and  {expanding the director field as} $\nv = n_1\ev_1+ n_2\ev_2+ n_3\bnu$, the components of $\Nv$ form  the  symmetric $3 \times 3$-matrix $(N_{ij} = n_i n_j)_{i,j=1,2,3}$.  Observe that $\nv$ and $\Nv$ are topologically different from  each other. In fact, the set of unit vectors  is in one-to-one correspondence  with the points of the two-dimensional unit sphere $\mathbb{S}^2$, while the set of second-order tensors in the form $\mathbf{a}\otimes\mathbf{a}$, with $|\mathbf{a}|=1$, is in bijective correspondence with the real projective $\mathbb{RP}^2$ plane which, as known, is diffeomorphic to the quotient space of $\mathbb{S}^2$ under the equivalence relation induced by the antipodal map. We also note that using $\Nv$ as a state variable for nematics is equivalent to fixing $s=1$ in the $\Qv-$tensor theory. 

We shall see that, given an energy density, taking the alignment tensor $\Nv$ as  the  state variable allows a richer variety of equilibrium textures  than on using the director $\nv$ as a macroscopic  descriptor of the orientation of nematic molecules.  To justify this assertion, consider   a spherical shell. Due to  the topology of the sphere, at equilibrium, ferromagnets (whose orientation is described by a unit vector)  form vortices with integer topological charges, while nematics (whose orientation is more appropriately described by a second-order tensor) may form also vortices with  half-integer topological charge. The difference in the possible topological charges of a vortex between the two theories stems from the different orientability of the state variables: arrows for ferromagnets, lines for nematics.    

We start by considering an ordinary 3D nematics confined in a thin layer.  In the bulk the energy due to the distortion of the molecular field is given by the celebrated Frank formula\cite{frank:1958}, while at the boundary   a surface energy term promotes a tangential alignment of the director field. As a result of a dimensionality reduction conducted in section I in the supplementary information,  such an energy density  approximates to   the effective surface energy density
\be\label{enfun}
w_s(\Nv, \grads \Nv) =\underbrace{ \frac{k}{4} |\grads \Nv|^2}_{\equiv \mathscr{E}_{ex}} + \frac{\varrho}{2} \Nv \cdot (\bnu \ot \bnu),
\en
where $k>0$ is the reduced elastic stiffness and $\varrho$ represents the anchoring strength. The sign of $\varrho$ determines the direction of the easy axis. The easy axis is normal to the shell if $\varrho$ is negative, tangential if $\varrho$ is positive.
Denoting $q_i (i=1,..,6)$  the components of $\Nv$ with respect to the local basis $\mathcal{B}$ in the Voigt notation (see section II in the supplementary information for details), we can decompose  the exchange energy density of a nematics as 
\begin{align}\label{decQ}
\mathscr{E}_{ex} = \underbrace{\frac k4(\eth_\gamma q_i)^2}_{\equiv \w0}+\underbrace{\frac k4\mathscr{H}_{ij}q_iq_j}_{\equiv \wa}+\underbrace{\frac k4\mathscr{D}_{ij}^{(\gamma)}\kappa_\gamma\mathcal{L}_{ij}^{(\gamma)}}_{\equiv \wdmi},
\end{align}
where $\gamma=1,2$, $i,j = 1,...,6$, and the symbol $\eth$ denotes the modified tangential derivative \cite{Sheka:2020}. 
Except for the different dimensions of the vector spaces which $\mathbf{q}=(q_1,...,q_6)$ and the unit magnetization vector belong to, the exchange energy density \eqref{decQ} is formally similar to the one for magnetic systems\cite{Sheka:2015}.

This decomposition allows to separate the intrinsic effects from the spurious effects due to the embedding of a 2D surface into a 3D Euclidean space. In fact, the  term $w_0$  in the exchange energy density \eqref{decQ} involves only covariant derivatives and hence represents the intrinsic part of the exchange interaction. This term is zero if and only if the alignment tensor is uniform in the tangent plane. The other two terms, instead, account also for the extrinsic curvature of the shell.

 Since $\mathscr{H}$ is a symmetric $6 \times 6$-matrix whose elements are functions of the principal curvatures only (see section II in the supplementary information), the energy term $\wa$ in \eqref{decQ} couples the extrinsic curvature with the alignment and induces anisotropic effects. At each point, $\wa$ reaches its absolute minimum when the texture is purely tangential and the director field is  aligned along the principal direction with the smallest modulus of curvature. To explain this point, observe that $\wa$ can be rewritten as  $\wa = k[\kappa_1^2(1 - q_2) + \kappa_2^2(1-q_1)]/2$, with $q_1,q_2\in\mathcal{T}=\{(u,v)\in\mathbb{R}^2:u,v\in[0,1],u+v\leq1\}$, and check that, depending on the magnitude of the moduli of the principal curvatures, it reaches its minimum in $\mathcal{T}$ for $q_1=1$ or $q_2=1$.  As a particular case, since any unit vector tangent to  a sphere is principal, at each point on a spherical shell $\wa$ attains its absolute minimum at any  tangential alignment. In any case, the curvature-induced anisotropy  term is activated whenever the surface is curved. It combines with anchoring potential, strengthening its effects if $\varrho>0$ and weakening them if $\varrho<0$.

Observe that the intrinsic and anisotropic terms are present also when considering purely  tangential  order parameters \cite{Napolia:2012, Napolib2012}, while the occurrence of the energy term $\wdmi$ is completely novel in the theory of nematic shells.  
This term  is responsible for a curvature-induced chiral interaction. The Lifshitz invariants of $\Nv$, $ \mathcal{L}_{ij}^{(\gamma)}= q_i\eth_\gamma q_j-q_j\eth_\gamma q_i $,  combines with the principal curvatures $\kappa_\gamma$, with  $\mathscr{D}^{(\gamma)}_{ij}$  being skew-symmetric matrices with constant entries, giving rise to an anisotropic DMI with geometry-dependent coefficients. A  term similar to $\wdmi$  occurs in the Landau-de Gennes theory of cholesteric LCs (which are intrinsically chiral materials) and is {  responsible for the emergence of  two-dimensional modulated states and the formation of  vortices or skyrmion lattices} \cite{Duzgun:2018, Duzgun:2021}. In the supplementary information we  prove that in terms of the components of the alignment tensor $\wdmi$ reads 
\be\label{wd}
\wdmi=k\;\eps_{ij\alpha}\eps_{\alpha\beta}L_{\beta\gamma}N_{ih}\eth_\gamma N_{hj},
\en
where $L_{\beta\gamma}$ are the components of the extrinsic curvature tensor.
Comparing \eqref{wd} with the chiral term in the energy density (1) in \cite{Duzgun:2018} we realize that the role played by  the (constant) helicity $q_0$ in cholesterics is here substituted by the second-order tensor $\boldsymbol{\tau}=\eps_{\alpha\beta}L_{\beta\gamma}\ev_\alpha\otimes\ev_\gamma$. This tensor can be regarded as a sort of helicity tensor which accounts for the anisotropy  induced by the extrinsic curvatures. The  components of $\boldsymbol{\tau}$ with respect to the local basis $\mathcal{B}$ form the $3\times3$-matrix
\be\label{tau}
[\boldsymbol{\tau}]=\left(\begin{array}{ccc}
0 & \kappa_2 &0\\
[3mm]
-\kappa_1 & 0 & 0\\
[3mm]
0 & 0 &0
\end{array}\right).
\en
From \eqref{tau} it is easy to deduce that  $\boldsymbol{\tau}$ is skew-symmetric on spheres, symmetric on minimal surfaces.

A direct inspection shows that the essential ingredient for the DMI, apart from curvature, is the existence of an out-of-plane component of the texture. The DMI  term favours rotation of the alignment around the normal direction, where the sense of rotation, in both nematic and magnetic shells,  is determined by the sign of the principal curvatures.  
 This term is instead suppressed in purely tangential or purely normal textures. The influence of surface geometry and easy axes on local interaction terms is sketched  in Figure 2 in the paper by Sheka et al. \cite{Sheka:2020}. 
 
 In spherical ferromagnetic shells, the interplay between the DMI term and the normal anisotropy ($\varrho<0$) is responsible of the emergence of skyrmions states as perturbations of  hedgehogs-like ground states \cite{Kravchuk:2016}. Very recently, LCs skyrmions have been realized as
micron sized solitons in a chiral nematic material confined between two parallel substrates \cite{Duzgun:21}.

Note that, purely tangential field are admissible on surfaces of genus one, while are not allowed on surface of genus zero. Thus, on a spherical shell, except for the hedgehog configuration,  all terms of the local interaction are unavoidably activated.

\begin{figure*}
\includegraphics[width=\linewidth]{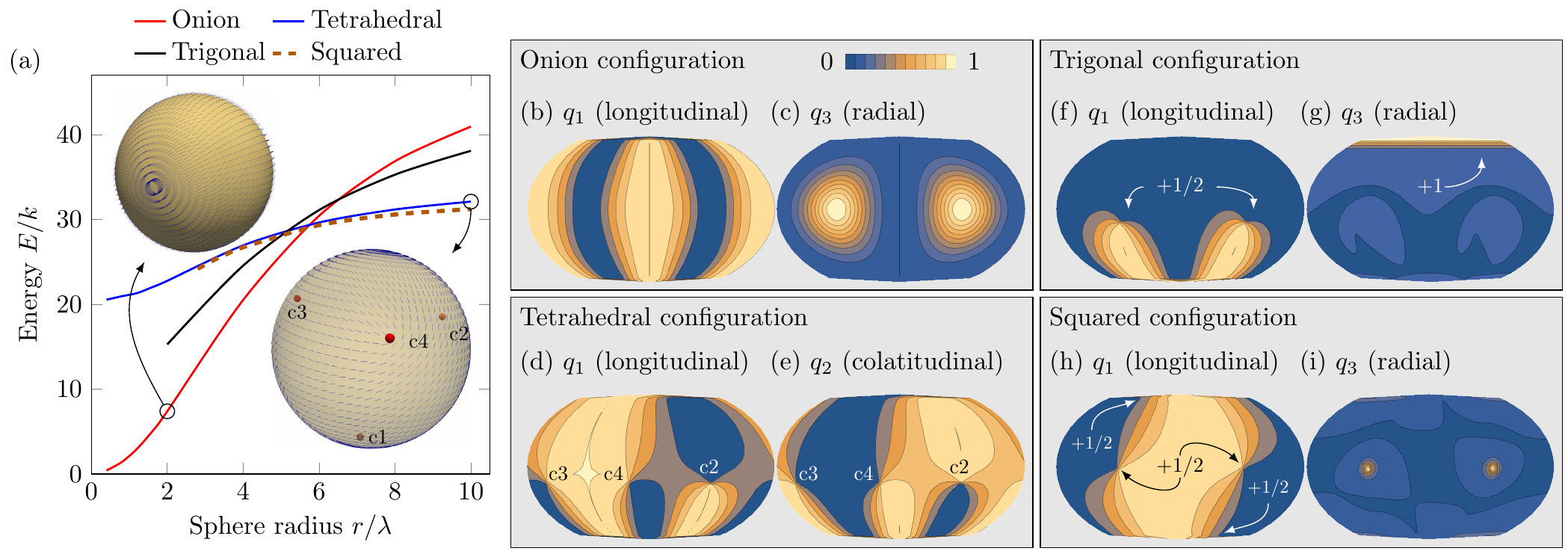}
\caption{\textbf{Equilibrium textures on nematic spherical shells.} (a) Energies of  {stable} onion, trigonal, tetrahedral  {and squared} configurations as functions of the ratio $r/\lambda$. Insets show the configurations of the nematic director field. Red spots mark the positions of disclinations of the tetrahedral texture.  Longitudinal ($q_1$) and radial ($q_3$) components of the onion configuration are represented by Kavrayskiy VII map projection in  (b,c), where  areas with different colours are separated by isolines.  Longitudinal ($q_1$) and colatitudinal ($q_2$) components of the tetrahedral configuration are displayed in (d,e).    {Longitudinal and radial components of trigonal and squared  configurations are sketched in (f,g,h,i). }}
\label{fig:simulations}
\end{figure*}

We now address the equilibrium textures on nematic  spherical shells by the numerical minimization of the energy functional \eqref{enfun} under appropriate constraints on the variables $q$'s (see  the Supplementary Information for details). We have used a  finite differences scheme  with  constant steps $\alpha = \pi/40$ for colatitude and longitude, and $\Delta r = 2\lambda$, with $\lambda =\sqrt{k/(2\varrho)}$ being a characteristic length which measures the
scale over which the alignment tensor varies on the shell at equilibrium, for the radius   $r\in [0.4\lambda, 10\lambda]$. The texture with the lowest energy  results from the interplay between the exchange and anchoring energies. For small enough values of the ratio $r/\lambda$ the exchange energy  dominates, which leads to axisymmetric textures that are almost uniform far from two small regions around two antipodal points where the  radial  component ($q_3$) is pronounced (Figures ~\ref{fig:simulations}(a,b,c)).   This configuration corresponds to the 3D  onion texture in ferromagnetic spherical shells\cite{Kong08a, Sloika17}. 

As the ratio $r/\lambda$ increases the role played by the anchoring energy becomes more and more important, the nematics is forced to align tangentially to the shell, and more complex textures occur. We find that for $r/\lambda\geq \xi^\star\approx 6$ the equilibrium texture  with four disclinations  has lower energy than the onion configuration (Figures ~\ref{fig:simulations}(a,d,e {,h,i})).  {In particular, we have found two stable  configurations with four +1/2 disclinations: one in which the disclinations are located at the vertices of a square inscribed in a great circle (Figures ~\ref{fig:simulations}(h,i)), and the other with disclinations situated  at the vertices of a tetrahedron (Figures ~\ref{fig:simulations}(d,e)). The former configuration (which, for brevity, we shall henceforth refer to as the squared configuration) has  slightly lower energy than the latter, which is instead stable for any $r/\lambda\in[0.4,10]$.}  {This result is in disagreement with other theoretical predictions in the literature on thin nematic shells \cite{Lubensky:1992, Vitelli:2006} in which it is claimed that the tetrahedral configuration is  the ground state on spherical shells of any radius. A possible experiment with particles of different sizes would clarify this controversy.    Instead, the metastability of these solutions is in agreement with experimental observations  \cite{Lopez-Leon11,Lopez-Leon:2011prl} and Monte Carlo simulations \cite{Bates:2008}. It should be noted, that theoretical models with the purely tangential alignment  predict that the configuration with two defects is unstable, in favour of the tetrahedral configuration \cite{Vitelli:2006,Napoli:2021}. Stable configurations with two disclinations are instead reported for  {thick spherical shells  or spherical shells with non-uniform thickness} \cite{Fernandez-Nieves:2007, Koning:2013}.}

A  {fourth}  {metastable} solution we have tested numerically is the trigonal configuration with three disclinations  located at the vertices of an isosceles triangle. One disclination has topological charge equal to $+1$, the other two equal to $+1/2$ (Figures ~\ref{fig:simulations}(f,g)). The distance between  {the two $+1/2$ disclinations} decreases as the ratio $r/\lambda$ decreases.  {The trigonal configuration has been experimentally observed and theoretically predicted  {(within the Frank theory)} in \cite{Lopez-Leon11}.}  {Unlike our findings, the theory proposed in \cite{Lopez-Leon11} explains  that the mutual distance between the disclinations of the trigonal configuration does not depend on the radius of the shell}.

 {Textures with a single  {+2 disclination}  or with two antipodal disclinations of charge +1/2 and +3/2 are configurations observed in cholesteric LCs shells \cite{Darmon:2016SM}. Despite the presence of the chiral term \eqref{wd} in the energy, we  found them unstable for any $r/\lambda\in [0.4, 10]$.}

Some of these shell textures computed on a sphere through our model approach agree with those
obtained using the Frank 2D theory \cite{Vitelli:2006}, Landau-de Gennes  2D theory \cite{Kralj:2011, Nitschke:2015}, Monte Carlo simulation studies \cite{Shin:2008, Bates:2010}.

Numerical results show that, although the energy density can be regarded as being the same in the two theories with $\nv$ and $\Nv$, the class of solutions admitted within the tensor theory is wider than that of vector theory. Thus, while ferromagnetics can admit only vortices of integer charge, nematics can also admit those of half-integer charge.  {This aspect has already been addressed in the literature of LCs. Indeed},  Ball and Zarnescu \cite{Ball:2008} discuss the case of a nematic in a planar configuration, confined in a finite plane region with two holes. Reporting  {their} words: 
``...it turns out that there are continuous, even smooth, line fields for which it is impossible to ‘assign arrows to the lines’; that is, one cannot make a choice of a vector out of each line in such a way that no discontinuity is created''.  {It is clear that  discontinuous configurations are not allowed  as they require  infinite energy}. 

When trying to force a field of vectors or lines on a surface $S$ with the topology of the sphere, discontinuities are unavoidable. 
Topological defects arise and, according to  the Poincar\'e-Hopf theorem, the sum of the topological charges of all defects on $S$ must be equal to 2, the Euler-Poincar\'e characteristics of a sphere. The topological charge of each defect is an integer multiple of ${1}/{2}$. However, it must be said that defects might not be the only points at which the solution lacks continuity. In fact, while it is always possible to have a continuous field of lines on the whole sphere (of course, excluding the points where the defects are located), the same is not true for fields of vectors. In particular, in the presence of a defect with fractional charge there are unavoidably curves through which the vector field is discontinuous. An example can be found in the work of Vitelli and Nelson \cite{Vitelli:2006} where such discontinuities are branch cuts in the complex plane. So, assuming   the points where the defects are excluded, there would be more continuous configurations allowed with lines (nematics) than with arrows (ferromagnets). When the field of lines or vectors can escape from the tangent plane, topological defects no longer exist. However, since out-of-plane escape has a certain energy cost, the texture tends to lie on the surface when possible. For topological reasons this is not possible at all points on a sphere and, therefore,  uplifts reverts to localized vortices that can have different structures depending on the theory. In other words, the texture around a lifting point mimics that of topological defects.

In conclusion, we have introduced a novel theory for nematic shells in which the optical axis is assumed not to be purely tangential. Accordingly, we have observed the occurrence of  novel curvature-induced effects that have been observed within theories for ferromagnetic shells. In particular,   the extrinsic curvature can be seen as a source of chirality  responsible for the onset of skyrmion- or meron-type nonlinear waves. On the other hand, it must be said that the difference in the orientability of the order parameters yields a class of equilibrium solutions for nematics wider than that for ferromagnets.

We believe that the impact of this work goes beyond the liquid crystals. In particular, we expect that the theory can be extended to curvilinear antiferromagnets. These are  be described by several vector order parameters. In the simplest case, the $\sigma$-model allows antiferromagnets to be described by the N\'{e}el vector\cite{Ivanov05a}. Being a direction by definition, in many cases  {the N\'eel vector} behaves  {like the nematic director and hence it allows} textures with fractional topological charges \cite{Dzyaloshinskii77,Ivanov01b,Kireev19,Galkina20}. The concept of  curvilinear antiferromagnetism has been introduced in \cite{Pylypovskyi20}. One can expect that curvilinear antiferromagnets admit novel textures similar to  {those predicted here for} nematics.

\section*{Acknowledgments}
D.~S.  acknowledges  the  financial  support  from the Ministry of Education and Science of Ukraine (project 19BF052-01).
The work of G.~N. and L.~V. has been funded by the MIUR (Italian Ministry of Education, University and Research) project PRIN 2017, ”Mathematics of active materials: From mechanobiology to smart devices”, project n. 2017KL4EF3.

\end{document}



\title{Supplementary Information: Nematic versus ferromagnetic shells: new insights in curvature-induced effects}

\author{Gaetano Napoli}
\email{gaetano.napoli@unisalento.it}
\affiliation{Dipartimento di Matematica e Fisica "E. De Giorgi", Universit\`a del Salento, Lecce (Italy).}
\author{Oleksandr V. Pylypovskyi}
\email{o.pylypovskyi@hzdr.de}
\affiliation{Helmholtz-Zentrum Dresden-Rossendorf e.V., Institute of Ion Beam Physics and Materials Research, 01328 Dresden, Germany}
\affiliation{Kyiv Academic University, 03142 Kyiv, Ukraine}
\author{Denis~D.~Sheka}
\email{sheka@knu.ua}
\affiliation{Taras Shevchenko National University of Kyiv, 01601 Kyiv, Ukraine}

\author{Luigi Vergori}
\email{luigi.vergori@unipg.it}
\affiliation{Dipartimento di Ingegneria, Universit\`a di Perugia, Perugia (Italy)}

\date{\today}

\begin{abstract}
In this supplementary material we derive the surface Frank-Rapini-Papoular  energy density  as a result of a perturbation  analysis of the usual 3D energy functional under the assumption that the region occupied by the nematics is very thin. We then decompose the energy density as the sum of the four contributions that are  discussed in details in the Letter. Finally, by means of the numerical minimization of the energy density we find  equilibrium textures on a spherical nematic shell with two, three and four disclinations. 
\end{abstract}

\maketitle

\section{Parametrization of the region occupied by the nematic shell. Differential operators}

  Let us assume that the nematic shell occupies a thin region $V$ of thickness $h$ around a regular compact surface $\esse$. Let $\bnus$ be the normal unit vector field to $\esse$. We parametrize points in the bulk through a coordinate set $(u,v,\xi)$ such that
\be\label{para}
p(u,v,\xi) =  p_S(u,v) + \xi \bnus (u,v),
\en
where $p_S$ is the normal projection of $p$ onto $\esse$, and  {$|\xi|$}, with $\xi\in[-h/2,h/2]$, is the distance of $p$ from the same surface. Such a coordinate set is well defined in a finite neighbourhood of $\esse$.  Next,   we introduce the principal curvatures $\kappa_{1s}(p_\esse)$ and $\kappa_{2s}(p_\esse)$ of $\esse$ at  $p_\esse\in \esse$, and  assume that
\be \label{hl}
h \ll \min_{p_\esse\in\esse}\left(\max\{|\kappa_{1s}(p_\esse)|, |\kappa_{2s}(p_\esse)|\}\right)^{-1} = \ell.
\en

For every fixed $\xi\in[-h/2,h/2]$, equation \eqref{para} defines a parallel surface $\esse_{\xi}=\{p_\esse+\xi\bnus(p_\esse):p_\esse\in\esse\}$ located at distance $|\xi|$ from $\esse$ with the unit normal vector field $\bnu:p\in\esse_\xi\mapsto \bnus(p_\esse)$. In this way, the unit vector field $\bnu$ is defined on the entire region $V$. The second-order tensor {$\grad \bnu$} is symmetric. Its eigenvectors  are  $\bnu$ (with a null eigenvalue)  and the unit vector fields
 $$\ev_i: p=p_\esse+\xi \bnus\in V\mapsto  \ev_{is}(p_\esse)\quad (i=1,2),$$
with $\ev_{1 s}$ and $\ev_{2 s}$ denoting the  principal directions fields  on $\esse$.    The spatial gradients of the  eigenvectors $\bnu$, $\ev_1$ and $\ev_2$ are\cite{Napolib2012}:
\begin{subequations} \label{eq:nabla-e-k}
\be \label{guno}
\nabla\bnu=-\frac{\c}{1-\xi \c}\ev_1\ot\ev_1-\frac{\cc}{1-\xi \cc}\ev_2\ot\ev_2,
\en
\be \label{gdue}
\nabla \ev_1=  -\ev_2\ot\bom+\frac{\c}{1-\xi \c}\bnu\ot\ev_1,
\en
\be \label{gtre}
\nabla \ev_2=\ev_1\ot\bom+\frac{\cc}{1-\xi \cc}\bnu\ot\ev_2,
\en
where  $\bom=-(\Omega_1\ev_1+\Omega_2\ev_2)$,
\be\label{chr1}
\left.\begin{array}{ll}
 \Omega_1=-\displaystyle\left[\frac{\Omega_{1s}}{1-\xi \kappa_{2s}}+\frac{\xi \grads \kappa_{1s}\cdot \ev_{2 s}}{(1-\xi \kappa_{1s})(1-\xi \kappa_{2s})}\right],\\
 [5mm]
 \Omega_2= -\displaystyle\left[\frac{\Omega_{2s}}{1-\xi \kappa_{1s}}-\frac{\xi \grads \kappa_{2s}\cdot \ev_{1 s}}{(1-\xi \kappa_{1s})(1-\xi \kappa_{2s})}\right],
 \end{array}\right.
\en
\end{subequations}
and $\Omega_{1s}$ and $\Omega_{2s}$ are the geodesic curvatures of the lines of curvature on $\esse$.

The differential operator $\grads$ in \eqref{chr1}  is the surface gradient. To  introduce this differential operator in the most general setting, let   $\Phi$ be a  smooth scalar, vector or tensor field defined on $\esse$.  Then the surface gradient of $\Phi$ is the tensor field resulting from the composition of the usual gradient of $\Phi$ and the projection onto the tangent plane of $\esse$, $\Pv=\mathbf{I}-\bnus\ot\bnus$, namely
\be\label{sgrad}
\grads \Phi= (\grad \Phi) \Pv.
\en
In analogy with the usual gradient operator, for a given vector field $\mathbf{u}$ we define the surface divergence and the surface curl of $\mathbf{u}$ as the trace of $\grads \mathbf{u}$ and twice the axial vector corresponding to  the skew-symmetric part of $\grads\mathbf{u}$, respectively.   In formulas we have 
$$
\dvs\mathbf{u}=\mathrm{tr}\grads\mathbf{u}=\grads\mathbf{u}\cdot \Pv,\quad\quad
\rots\mathbf{u}=-\veps\grads\mathbf{u},
$$
where $\veps$ denotes the Ricci alternator.

\section{Derivation of the surface Frank-Rapini-Papoular energy}

Let  $\nv$ be a unit vector, called the director, which represents the average alignment of the molecules of the nematics.   The celebrated Frank formula\cite{frank:1958} for the elastic energy density (per unit of volume) associated with the director distortion consists of four terms
\begin{align}\label{OZF}
2w_{OZF} & = K_1 (\dv \nv)^2 + {K_2}(\nv \cdot \rot \nv)^2 + {K_3}|\nv \times \rot \nv|^2 \nonumber \\ &+ (K_2 + K_{24})[(\nabla \nv)\cdot(\nabla \nv)^T - (\dv \nv)^2]
\end{align}
where the	constants	$K_1$, $K_2$, $K_3$, and $K_{24}$ are called the splay, twist, bend, and saddle-splay moduli, respectively. To ensure a stable undistorted configuration of a nematic liquid crystal in the absence of external fields or confinements, the three moduli $K_i (i = 1, 2, 3)$ must be non-negative, whereas the elastic saddle-splay constant must obey Ericksen's inequalities\cite{ericksen:1966}:
$$
|K_{24}|\leq K_2, \qquad	K_2 +K_{24}\leq 2 K_1.
$$

In  the attempt to capture also the effects of the boundaries $\partial \esse_{\pm h/2}$ on the orientation of the nematic director, we consider the  energy functional resulting from the sum of the Frank energy and the Rapini-Papoular surface anchoring energy\cite{Rapini69}
\begin{align}\label{energy}
\nonumber
W &= \int_{V} w_{OZF}(\nv, \grad \nv)  \d V\\
&+\frac \varrho4\int_{\esse_{h/2}\cup \esse_{-h/2}}(\nv\cdot\upbv)^2\d a,
\end{align}
where the constant $\varrho$ represents the anchoring strength and $\upbv$ is the unit outward normal to $\esse_{\pm h/2}$. The  anchoring energy in \eqref{energy}   penalizes the deviation of the molecules at the boundaries $\esse_{\pm h/2}$ from tangential directions if $\varrho$ is positive, from the normal direction if $\varrho$ is negative.

 Under the assumption \eqref{hl} on the smallness of the thickness of the region $V$ occupied by the nematics, the energy functional \eqref{energy} approximates to a surface integral over $S$, with the integrand representing  the surface Frank-Rapini-Papular energy density. To prove this, we first assume that the director field    $\nv$  does not vary along directions normal to $\esse$, that is it satisfies the property
\be\label{nprop}
\nv(p)= \nv(p_S)
\en
 at each point $p=p_S+\xi\bnus\in V$.   We then  parametrize the director field as
\be\label{parn}
\nv=\sin\theta\cos\varphi\ev_1+\sin\theta\sin\varphi\ev_2+\cos\theta\bnu,
\en
where, in view  of ansatz \eqref{nprop}, the angles $\theta$ and $\varphi$ satisfy the properties
\be\label{props}
\theta(p)=\theta(p_S) \quad \textrm{and} \quad \varphi(p)=\varphi(p_S)
\en
for all $p=p_S+\xi\bnus\in V$.

From \eqref{guno}-\eqref{gtre} and \eqref{parn}  the gradient of $\nv$ is found to be
\begin{align}\label{gradn}
 \nabla\nv=(\bnu\times\nv)\otimes(\grad \varphi-\bom)-\bnu\otimes(\nabla\bnu)\nv  \nonumber \\
 +(\bnu\cdot\nv)\nabla\bnu-\displaystyle \frac{\bnu-(\bnu\cdot\nv)\nv}{|\bnu\times\nv|}\otimes\grad \theta,
\end{align}
where, due to \eqref{props} and following the similar arguments as  in Appendix A in \cite{Napolib2012},
\be\label{th}
\grad \theta =\left(\frac{\grads \thetas\cdot\evus}{1-\xi \c}\right)\ev_1+\left(\frac{\grads \thetas\cdot\evds}{1-\xi \cc}\right)\ev_2,
\en
and
\be\label{vp}
\grad \varphi =\left(\frac{\grads \varphis\cdot\evus}{1-\xi \c}\right)\ev_1+\left(\frac{\grads \varphis\cdot\evds}{1-\xi \cc}\right)\ev_2.
\en
 The scalar fields $\thetas$ and $\varphis$ on the right-hand sides of equations \eqref{th} and \eqref{vp} are  the restrictions of  $\theta$ and $\varphi$  to the surface $S$, namely $\thetas:p_\esse\in\esse\mapsto\theta(p_\esse)$ and $\varphis:p_\esse\in\esse\mapsto\varphi(p_\esse)$.
 From \eqref{guno}, \eqref{chr1} and \eqref{gradn}--\eqref{vp} it is evident that although the director field is assumed to be constant along directions normal to $\esse$, its spatial gradient \emph{does} depend on the spatial variable $\xi$ in the normal direction $\bnu$.

At $\xi=0$ the parametrization of $\nv$ \eqref{parn} and \eqref{props} give the restriction $\nvs$ of the director field to the surface $\esse$,
\be\label{nesse}
\nvs=\sin\thetas\cos\varphis\evus+\sin\thetas\sin\varphis\evds+\cos\thetas\bnus,
\en
while evaluating  \eqref{guno}-\eqref{chr1}, \eqref{th} and \eqref{vp}  at $\xi=0$ gives the surface gradient of   $\nvs$,
\ba\label{gradsn}
 \grads\nvs=(\bnus\times\nvs)\otimes(\grads \varphis-\boms)+\bnus\otimes\Lv\nvs  \nonumber \\ -(\bnus\cdot\nvs)\Lv   -\displaystyle \frac{\bnus-(\bnus\cdot\nvs)\nvs}{|\bnus\times\nvs|}\otimes\grads \thetas,
\ea
where
\begin{subequations}
\be
\boms=-\Omega_{1s}\evus-\Omega_{2s}\evds, 
\en
\be
\Lv=\c\evus\ot\evus+\cc\evds\ot\evds,
\en
\end{subequations}
are the vector parametrizing the spin connection on $\esse$, and the extrinsic curvature tensor of $\esse$, respectively. Incidentally, we recall that the mean and Gaussian curvatures of $\esse$ are defined as
\be
H=\frac12\mathrm{tr}\Lv \quad \textrm{and} \quad K=\frac12\left[(\mathrm{tr}\Lv)^2-\mathrm{tr}(\Lv^2)\right],
\en
 respectively.

Recalling the definition of the parameter $\varepsilon$ (c.f. \eqref{hl}), from \eqref{gradn}--\eqref{gradsn} we deduce that (see \citet{Napolib2012} for details)
\be\label{approx}
\grad \nv = \grads \nvs + o(\epsilon) \quad \textrm{as } \varepsilon\rightarrow0,
\en
whence
\be\label{approx2}
\dv \nv= \dvs \nvs + o(\epsilon), \quad \rot \nv=\rots\nvs+o(\epsilon)
\en
as $\varepsilon\rightarrow0$. 

With the parametrization \eqref{para} the volume element $\d V$ reads
\be\label{vel}
\d V =(1-2H\xi+K\xi^2)\d A \d\xi
\en
where $\d A$ is the area element on $\esse$. On the other hand, 
on the boundaries $S_{\pm h/2}$ the unit outward normal and the area element are $\upbv=\pm\bnu_s$ and
\be\label{ael}
\d A_{\pm h/2}=\left(1\mp hH +\frac{h^2}{4}K\right)\d A,
\en
respectively.
In view of \eqref{hl} the volume and area elements\eqref{vel} and \eqref{ael} approximate to
\be\label{var}
\d V=\d A \d\xi +o(\eps) \quad \textrm{and} \quad \d A_{ \pm h /2}= \d A +o(\eps) 
\en
as  $\eps\rightarrow 0$. 

Finally, from \eqref{nprop}, \eqref{nesse}, \eqref{approx}, \eqref{approx2}, \eqref{ael} and \eqref{var}  we deduce that
\begin{widetext}
\begin{align}\label{2df0}
\nonumber
W&=\frac12\int_V\left\{K_1(\dv\nv)^2+K_2(\nv\cdot\rot\nv)^2+K_3|\nv\times\rot\nv|^2+(K_2+K_{24})\left[(\grad \nv)\cdot(\grad \nv)^T-(\dv\nv)^2\right]\right\}\d V\\
\nonumber
&\quad+\int_{\esse_{h/2}}\frac\varrho4(\nv\cdot\bnu_s)^2\d A_{h/2}+\int_{\esse_{-h/2}}\frac\varrho4[\nv\cdot(-\bnu_s)]^2\d A_{-h/2}\\
\nonumber
&=\frac12\int_{-h/2}^{h/2}\Bigg\{\int_\esse\left[K_1(\dv\nv)^2+K_2(\nv\cdot\rot\nv)^2+K_3|\nv\times\rot\nv|^2\right](1-2H\xi+K\xi^2)\d A\Bigg\}\d \xi\\
\nonumber
&\quad +\frac12\int_{-h/2}^{h/2}\Bigg\{\int_\esse(K_2+K_{24})\left[(\grad \nv)\cdot(\grad \nv)^T-(\dv\nv)^2\right](1-2H\xi+K\xi^2)\d A\Bigg\}\d \xi\\
\nonumber
&\quad+\int_{\esse}\frac\varrho4(\nv\cdot\bnu_s)^2\left(1- hH +\frac{h^2}{4}K\right)\d A +\int_{\esse}\frac\varrho4(\nv\cdot\bnu_s)^2\left(1+ hH +\frac{h^2}{4}K\right)\d A\\
\nonumber
&=\frac12\int_{-h/2}^{h/2}\Bigg\{\int_\esse\left[K_1(\dvs\nvs)^2+K_2(\nvs\cdot\rots\nvs)^2+K_3|\nvs\times\rots\nvs|^2\right]\d A\Bigg\}\d \xi\\
\nonumber
&\quad+\frac12\int_{-h/2}^{h/2}\Bigg\{\int_\esse(K_2+K_{24})\left[(\grads \nvs)\cdot(\grads \nvs)^T-(\dvs\nvs)^2\right]\d A\Bigg\}\d \xi+\int_{\esse}\frac\varrho2(\nvs\cdot\bnu_s)^2\d A+o(\eps)\\
\nonumber
&=\frac h2\int_\esse\left[K_1(\dvs\nvs)^2+K_2(\nvs\cdot\rots\nvs)^2+K_3|\nvs\times\rots\nvs|^2\right]\d A\\
&\quad +\frac12\int_\esse\left\{h(K_2+K_{24})[(\grads\nvs)\cdot(\grads\nvs)^T-(\dvs\nvs)^2]+\varrho(\nvs\cdot\bnu_s)^2\right\}\d A+o(\eps) \quad \quad \textrm{as } \eps\rightarrow0.
\end{align}
\end{widetext}
We have then proven that $W\approx W_s=\displaystyle\int_\esse w_s\d A$, with
\begin{align}\label{2df}
\nonumber
2w_s&=k_1(\dvs\nvs)^2+k_2(\nvs\cdot\rots\nvs)^2+k_3|\nvs\times\rots\nvs|^2\\
\nonumber
&+(k_2+k_{24})[(\grads\nvs)\cdot(\grads\nvs)^T-(\dvs\nvs)^2]\\
&+\varrho(\nvs\cdot\bnu_s)^2,
\end{align}
where $k_i=hK_i$ ($i=1,2,3,24$) are, respectively, the re-scaled splay, twist, bend and saddle-splay constants with the same physical dimensions as energy per unit area.

Within the one-constant approximation ($K_1=K_2=K_3=K$ and $K_{24}=0$ in \eqref{OZF}) the surface energy density \eqref{2df} reduces to
\be
w_s=\frac k2|\grads\nvs|^2+\frac\varrho2(\nvs\cdot\bnus)^2,
\label{frank2d}
\en
with $k=hK$. 

\section{Surface energy density in terms of the alignment tensor }

In what follows we shall  consider the surface energy density \eqref{frank2d}, and, for simplicity of notation, we shall omit the subscript $s$ to $\nv$, $\ev_1$, $\ev_2$ and $\bnu$.

We now show that the moduli of the surface gradients of the director $\nv$ and alignment tensor $\Nv=\nv\otimes\nv$ are  related each other. 

From the definition of surface gradient \eqref{sgrad}, the  components of the surface gradient of the alignment tensor  are found to be
\begin{align}\label{gradsN}
\nonumber
(\grads \Nv)_{ijh} &= \frac{\partial N_{ij}}{\partial x_l}P_{lh}= \frac{\partial n_i}{\partial x_l}P_{lh}n_j+n_i\frac{\partial n_j}{\partial x_l}P_{lh}\\
&\equiv n_{i;h}n_j+n_in_{j;h}.
\end{align}
(In \eqref{gradsN}, and in the subsequent formulas, the Einstein summation convention is used.) Since  $n_i^2=1$ (the director is a unit vector), we deduce that
$$
n_in_{i;h}=(n_i^2)_{;h}-n_in_{i;h}=-n_in_{i;h}\Longrightarrow n_in_{i;h}=0,
$$
which combined with \eqref{gradsN}  yields
\begin{align}\label{Nn}
\nonumber
|\grads\Nv|^2&=N_{ijh}^2=n_{i;h}^2n_j^2+2n_in_{i;h}n_jn_{j;h}+n_i^2n_{j;h}^2\\
&=2n_{i;h}^2=2\left(\frac{\partial n_i}{\partial x_l}P_{lh}\right)^2=2|\grads\nv|^2.
\end{align}
In view of this identity, the surface energy density \eqref{frank2d} can be rewritten in terms of the alignment tensor as
\be
w_s=\frac k4|\grads\Nv|^2+\frac\varrho2\Nv\cdot(\bnus\otimes\bnus),
\en

On the other hand,  with respect to the local  basis $\{\ev_1,\ev_2,\ev_3=\bnu\}$ on $\esse$,  the surface gradient of the alignment tensor reads
\be
\grads \Nv =[\grads\Nv\cdot(\ev_i\otimes\ev_j\otimes\ev_\gamma)]\,(\ev_i\otimes\ev_j\otimes\ev_\gamma),
\en
where,   Latin indices range from 1 to 3, Greek ones from 1 to 2 ( $\grads\Nv\cdot(\ev_i\otimes\ev_j\otimes\bnu)=0$ for all $i,j=1,2,3$), and
\begin{align}
\grads\Nv&\cdot(\ev_i\otimes\ev_j\otimes\ev_\gamma)=\nabla_\gamma\Nv\cdot (\ev_i\otimes\ev_j)\\
\nonumber
&=\nabla_\gamma N_{ij}+N_{hj}(\nabla_\gamma\ev_h)\cdot\ev_i+N_{ik}(\nabla_\gamma\ev_k)\cdot\ev_j.
\end{align}

In view of the symmetry of $\Nv$ we have  that
\be
\nabla_\gamma\Nv\cdot (\ev_i\otimes\ev_j)=\nabla_\gamma\Nv\cdot (\ev_j\otimes\ev_i),
\en
whence, with the aid of \eqref{eq:nabla-e-k} evaluated at  $\xi=0$, we obtain
\begin{subequations}\label{compgN}
\be
\nabla_\gamma\Nv\cdot (\ev_\alpha\otimes\ev_\beta)=\eth_\gamma N_{\alpha\beta}-(N_{3\beta}\delta_{\alpha\gamma}+N_{\alpha3}\delta_{\beta\gamma})\kappa_\gamma,
\en
\be
\nabla_\gamma\Nv \cdot(\bnu\otimes\ev_\beta)=\eth_\gamma N_{3\beta}+\kappa_\gamma(N_{\gamma\beta}-N_{33}\delta_{\beta\gamma}),
\en
\be
\nabla_\gamma\Nv \cdot(\bnu\otimes\bnu) = \eth_\gamma N_{33}+2N_{\gamma3}\kappa_\gamma,
\en
\end{subequations}
where
\begin{subequations}\label{ethQ}
\be
\eth_\gamma N_{\alpha\beta} =\nabla_\gamma N_{\alpha\beta}+(\varepsilon_{\alpha\zeta} N_{\zeta\beta}+\varepsilon_{\beta\zeta}N_{\alpha\zeta})\Omega_\gamma,
\en
\be
\eth_\gamma N_{3\beta}=\nabla_\gamma N_{3\beta}+\varepsilon_{\beta\zeta}N_{3\zeta}\Omega_\gamma,
\en
\be
\eth_\gamma N_{33} = \nabla_\gamma N_{33}.
\en
\end{subequations}

As known, the space of  symmetric second-order tensors is isomorphic to the six-dimensional Euclidean vector space. Specifically, on setting
\begin{equation} \label{voigt}
\left.\begin{array}{cc}
N_{11}=q_1, \quad N_{22}=q_2,  \quad N_{33}=q_3, \\
[5mm]
 \sqrt{2}N_{23}=q_4, \quad \sqrt{2}N_{13}=q_5, \quad \sqrt{2}N_{12}=q_6,
\end{array}\right.
\end{equation}
we can identify $\Nv$ with the vector $\mathbf{q}=(q_1,q_2,q_3,q_4,q_5,q_6)$. Relations \eqref{voigt} are known as the Voigt notation. On using  the Voigt notation it can be proven   that the space of second-order tensors and the six-dimensional Euclidean vector space are also isometric, whence $|\Nv|^2=|\qv|^2$, $|\grads\Nv|^2=|\grads \qv|^2$ and $(\eth_\gamma N_{ij})^2=(\eth_\gamma q_k)^2$.

 From \eqref{compgN}, \eqref{ethQ} and \eqref{voigt} we obtain (after lengthy algebra)
\begin{align}\label{decQ}
|\grads\Nv|^2&=[\nabla_\gamma N\cdot(\ev_i\otimes\ev_j)]^2\\
\nonumber&=(\eth_\gamma q_i)^2+\mathscr{H}_{ij}q_iq_j+\mathscr{D}_{ij}^{(\gamma)}\kappa_\gamma\mathcal{L}_{ij}^{(\gamma)},
\end{align}
where
\begin{subequations}
\be
\eth_\gamma q_1 = \nabla_\gamma q_1 +\sqrt{2}\Omega_\gamma q_6,
\en
\be
\eth_\gamma q_2 = \nabla_\gamma q_2-\sqrt{2}\Omega_\gamma q_6,
\en
\be
\eth_\gamma q_3 = \nabla_\gamma q_3,
\en
\be
\eth_\gamma q_4 = \nabla_\gamma q_4 -\Omega_\gamma q_5,
\en
\be
\eth_\gamma q_5 = \nabla_\gamma q_5+\Omega_\gamma q_4,
\en
\be
\eth_\gamma q_6 = \nabla_\gamma q_6-\sqrt{2}\Omega_\gamma(q_1-q_2),
\en
\end{subequations}
\be
\mathcal{L}_{ij}^{(\gamma)}=(q_i\eth_\gamma q_j-q_j\eth_\gamma q_i),
\en
 $\mathscr{H}=(\mathscr{H}_{ij})_{i,j=1,...,6}$  is the symmetric sixth order matrix
 \begin{widetext}
\begin{equation} \label{eq:Dijk-Q}
\begin{aligned}
\!\boldsymbol{\mathscr{H}}=\left(\begin{array}{cccccc}
2\kappa_1^2 & 0 & -2\kappa_1^2 & 0 & 0 & 0  \\
[3mm]
\displaystyle 0 & 2\kappa_2^2  &-2\kappa_2^2  & 0 & 0 & 0 \\
[3mm]
\displaystyle  -2\kappa_1^2 & -2\kappa_2^2 & 2(\kappa_1^2+\kappa_2^2)  & 0 & 0  & 0        \\
[3mm]
\displaystyle 0  &    0 &  0 &  \kappa_1^2+4\kappa_2^2 & 0 &   0                                \\
[3mm]
\displaystyle 0   & 0 &  0 & 0   & 4\kappa_1^2+\kappa_2^2  &0                                \\
[3mm]
\displaystyle  0 & 0 &0 & 0 & 0 &\kappa_1^2+\kappa_2^2
\end{array}\right),\!\!,\!\!
\end{aligned}
\end{equation}
\end{widetext}
and $\mathscr{D}^{(\gamma)}=(\mathscr{D}^{(\gamma)}_{ij})_{i,j=1,...,6}$ ($\gamma=1,2$) are skew-symmetric matrices the independent non-zero entries of which are
\begin{subequations}
\be
\mathscr{D}_{15}^{(1)}=-\mathscr{D}_{35}^{(1)}=\sqrt{2}, \quad  \mathscr{D}_{46}^{(1)}=-1, 
\en
and
\be
\mathscr{D}_{24}^{(2)}=-\mathscr{D}_{34}^{(2)}=\sqrt{2}, \quad \mathscr{D}_{56}^{(2)}=-1.
\en
\end{subequations}

Finally, in view of \eqref{Nn}, on using the Voigt notation the surface energy density  \eqref{frank2d} becomes
\be\label{enlett}
w_s=\underbrace{\frac k4 (\eth_\gamma q_i)^2}_{=\mathscr{E}_0}+\underbrace{\frac k4\mathscr{H}_{ij}q_iq_j}_{=\mathscr{E}_\mathrm{A}}+\underbrace{\frac k4 \mathscr{D}_{ij}^{(\gamma)}\kappa_\gamma\mathcal{L}_{ij}^{(\gamma)}}_{=\mathscr{E}_\mathrm{DM}} + \frac \varrho2 q_3.
\en

To find stable equilibrium configurations one has to minimize the energy functional $W_s=\int_\esse w_s\d A$ subject to some constraints on  the scalar fields $q$'s. In fact, since $q_1,...,q_6$ are  the components of the alignment tensor $\Nv=\nv\otimes\nv$ they must satisfy the following inequalities and equations
\begin{subequations}\label{constr}
\be\label{pos}
q_i\geq0 \quad  (i=1,2,3),
\en
\be\label{traceq}
q_1+q_2+q_3=1,
\en
\be\label{rank}
q_2q_3-\frac{q_4^2}{2}=q_1q_3-\frac{q_5^2}{2}=q_1q_2-\frac{q_6^2}{2}=0,
\en
\end{subequations}
Inequalities \eqref{pos} impose the non-negativeness of $q_1$, $q_2$ and $q_3$  as these components of $\Nv$ are the squares of the components of the nematic director $\nv$. Since $\nv$ is  a unit vector, the trace of $\Nv$ must be equal to unity, which justifies \eqref{traceq}. Finally, equations \eqref{rank} together with \eqref{pos} and \eqref{traceq} imply that  the components of $\Nv$ form a matrix with rank  equal to one as it has to be because the alignment tensor is of the form $\nv\otimes\nv$.

We conclude this section by observing that by using \eqref{compgN} the energy density  term $\mathscr{E}_\mathrm{DM}$ in \eqref{enlett} can be written also as
\be
\mathscr{E}_\mathrm{DM}=k\ricci_{ij\alpha}\ricci_{\alpha\beta}L_{\beta\gamma}N_{ik}\eth_\gamma N_{kj}.
\en
This expression allows a comparison between the Dzyaloshinskii–Moriya  interaction term $\mathscr{E}_\mathrm{DM}$ and the chiral term in the energy density (1) in \cite{Duzgun:2018} (see the Letter).

\section{Surface energy density on a spherical shell}

On a spherical shell of radius $r$ the local basis formed by the principal directions and the unit normal is $\{\ev_\varphi,\ev_\theta,-\ev_r\}$, with $\ev_r$, $\ev_\theta$ and $\ev_\varphi$ being the radial, co-latitudinal and longitudinal directions, respectively, the principal curvatures are $\kappa_\varphi=\kappa_\theta=1/r$ and the vector parametrizing the spin connection is $\boldsymbol{\Omega}=(\cot\theta/r)\ev_\varphi$. Therefore, for a spherical shell the energy density terms $\mathscr{E}_0$, $\mathscr{E}_\mathrm{A}$ and $\mathscr{E}_\mathrm{DM}$, and the whole energy density  \eqref{enlett} read, respectively,
\begin{align*}
\mathscr{E}_0=\frac{k}{4r^2}\Bigg\{&\left(q_{1,\varphi}\csc\theta +\sqrt{2}q_6\cot\theta\right)^2+q_{1,\theta}^2\\
\nonumber
&+\left(q_{2,\varphi}\csc\theta -\sqrt{2}q_6\cot\theta\right)^2+q_{2,\theta}^2\\
\nonumber
&+q^2_{3,\varphi}\csc^2\theta+q_{3,\theta}^2+\left(q_{4,\varphi}\csc\theta-q_5\cot\theta\right)^2\\
\nonumber
&+q^2_{4,\theta}+\left(q_{5,\varphi}\csc\theta+q_4\cot\theta\right)^2+q^2_{5,\theta}\\
\nonumber
&+\left[q_{6,\varphi}\csc\theta-\sqrt{2}(q_1-q_2)\cot\theta\right]^2+q_{6,\theta}^2\Bigg\},
\end{align*}
\begin{align*}
\mathscr{E}_\mathrm{A}=\frac{k}{4r^2}\Big(&2q_1^2+2q_2^2+4q_3^2+5q_4^2+5q_5^2+2q_6^2\\
\nonumber
&-4q_1q_3-4q_2q_3\Big),
\end{align*}
\begin{align*}
\mathscr{E}_\mathrm{DM}=\frac{k}{2r^2}&\Bigg\{\Big[\sqrt{2}(q_1q_{5,\varphi}-q_5q_{1,\varphi}-q_3q_{5,\varphi}+q_5q_{3,\varphi})\\
&\quad-q_4q_{6,\varphi}+q_6q_{4,\varphi}\Big]\csc\theta\\
&+\Big[\sqrt{2}q_4(2q_1-q_2-q_3)-3q_5q_6\Big]\cot\theta\\
&+\sqrt{2}(q_2q_{4,\theta}+q_4q_{3,\theta}-q_4q_{2,\theta}-q_3q_{4,\theta})\\
\nonumber
&-q_5q_{6,\theta}+q_6q_{5,\theta}\Bigg\},
\end{align*}
and
\begin{align}\label{eq:frank-curv}
\nonumber
w_s &= \frac{k}{4r^2} \Big[({  q_{1,\theta}}) ^2+\left(q_{1,\varphi}  \csc \theta -\sqrt{2} q_5 -  +\sqrt{2}  {  q_6}   \cot \theta)\right)^2 \\
 \nonumber
 &+\left({  q_{6,\varphi}}  \csc \theta-{  q_4} -   \sqrt{2}  {  q_1}   \cot \theta+\sqrt{2}  {  q_2}   \cot \theta\right)^2\\
 \nonumber
 &+\left(  {  q_{5,\varphi}} \csc \theta  +\sqrt{2}  {  q_1}-\sqrt{2}  {  q_3}+ {  q_4}   \cot \theta\right)^2 \\
 \nonumber
 &+\left( {  q_{2,\theta}} -\sqrt{2}  {  q_4}\right)^2+\left(   {  q_{2,\varphi}}  \csc \theta -\sqrt{2}  {  q_6}   \cot \theta\right)^2\\
 \nonumber
 &+\left( {  q_{4,\theta}} +\sqrt{2}  {  q_2}-\sqrt{2}  {  q_3}\right)^2+\left( {  q_{3,\theta}} +\sqrt{2}  {  q_4}\right)^2\\
 \nonumber
 &+\left(  {  q_{3,\varphi}}  \csc \theta +\sqrt{2}  {  q_5}\right)^2 
 +\left(  {  q_{4,\varphi}} \csc \theta  - {  q_5}   \cot \theta+ {  q_6}\right)^2\\
 &+\left( {  q_{5,\theta}} + {  q_6}\right)^2 +\left(     q_{6,\theta}- q_5 \right)^2\Big] + \frac{\varrho}{2} q_3.
\end{align}

\section{Simulations}

To determine the equilibrium configuration of nematic molecules, we numerically minimize the energy functional $W_s$ with energy density  as  in \eqref{eq:frank-curv}. To do this, the continual fields $q_l$, $l = \overline{1,6}$ are replaced by their discrete counterparts on a finite difference mesh (FDM) with constant steps along the longitudinal and colatitudinal directions $\Delta\phi = 2\pi/M$ and $\Delta\theta=\pi/N$, respectively, where $N > 1$ and $M > 1$ are integer numbers determining discretization of the mesh. If $f^{ij}$ represents the FDM counterpart of the function $f(\theta,\phi)$ on a sphere, then
\begin{equation}
\begin{aligned}
f(i\Delta\theta, j \Delta\phi) &= f^{ij},\\
f(\theta = 0) &= f^0,\quad f(\theta = \pi) = f^N,\\
i&=\overline{1,N-1},\quad j=\overline{0,M-1}.
\end{aligned}
\end{equation} 
Integration of $W_s$ is performed using the double Riemann sum method. We minimize the function $W_\text{tot} = W_s + W_p$ using the conjugate gradients method with the penalty function

\begin{widetext}
\begin{equation}\label{eq:penalty}
\begin{aligned}
W_p & = \Lambda\sum_{ij} \left\{ (1-q_1^{ij}-q_2^{ij}-q_3^{ij})^2 + \left[2q_1^{ij}q_2^{ij}-\left(q_6^{ij}\right)^2\right]^2 + \left[2q_2^{ij}q_3^{ij}-\left(q_4^{ij}\right)^2\right]^2 + \left[2q_3^{ij}q_1^{ij}-\left(q_5^{ij}\right)^2\right]^2 \right\}\\
& + \Lambda \left\{ (1-q_1^0-q_2^0-q_3^0)^2 + \left[2q_1^0q_2^0-\left(q_6^0\right)^2\right]^2 + \left[2q_2^0q_3^0-\left(q_4^0\right)^2\right]^2 + \left[2q_3^0q_1^0-\left(q_5^0\right)^2\right]^2 \right.\\
& \left. + (1-q_1^0-q_2^0-q_3^0)^2 + \left[2q_1^Nq_2^N-\left(q_6^N\right)^2\right]^2 + \left[2q_2^Nq_3^N-\left(q_4^N\right)^2\right]^2 + \left[2q_3^Nq_1^N-\left(q_5^N\right)^2\right]^2 \right\},
\end{aligned}
\end{equation} 
\end{widetext}
where $\Lambda = 10^3$. To avoid saddle points of $W_\text{tot}$, the multiplier $q_3$ is replaced by $1 - q_1 - q_2$ in the anchoring term.

\begin{figure*}
\includegraphics[width=\linewidth]{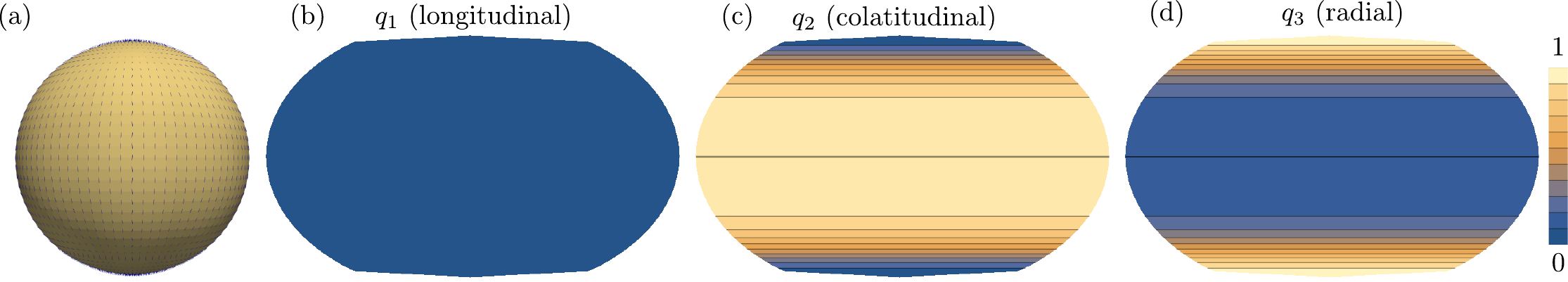}
\caption{\textbf{Onion configuration of the nematic director on the sphere of radius $r = 4\lambda$ ($\lambda =\sqrt{k/(2\varrho)}$.} (a) Flux lines of the director field on the sphere. (b--d) Components of alignment{tensor} tensor $\Nv$.}
\label{fig:supp-onion}
\end{figure*}

To test stability of different configurations, the strictly in-surface initial state is given 
\begin{equation}\label{eq:ini-state}
\begin{aligned}
q_1 & = \cos^2 \sigma,\quad q_2 = \sin^2 \sigma,\quad q_6 = \sqrt{2}\sin\sigma \cos\sigma,\\
q_3 & = q_4 = q_5 = 0
\end{aligned}
\end{equation}
for the sphere of each radius with the concrete configuration determined by the angle $\sigma$. For the onion configuration it reads
\begin{equation}\label{eq:onion-ini}
\sigma_\text{onion} = \pi/2.
\end{equation}
The relaxed state is shown in Fig.~\ref{fig:supp-onion}. Note, that in general case, the axis of symmetry not necessary to coincide with the direction along the sphere poles. 

\begin{figure*}
\includegraphics[width=\linewidth]{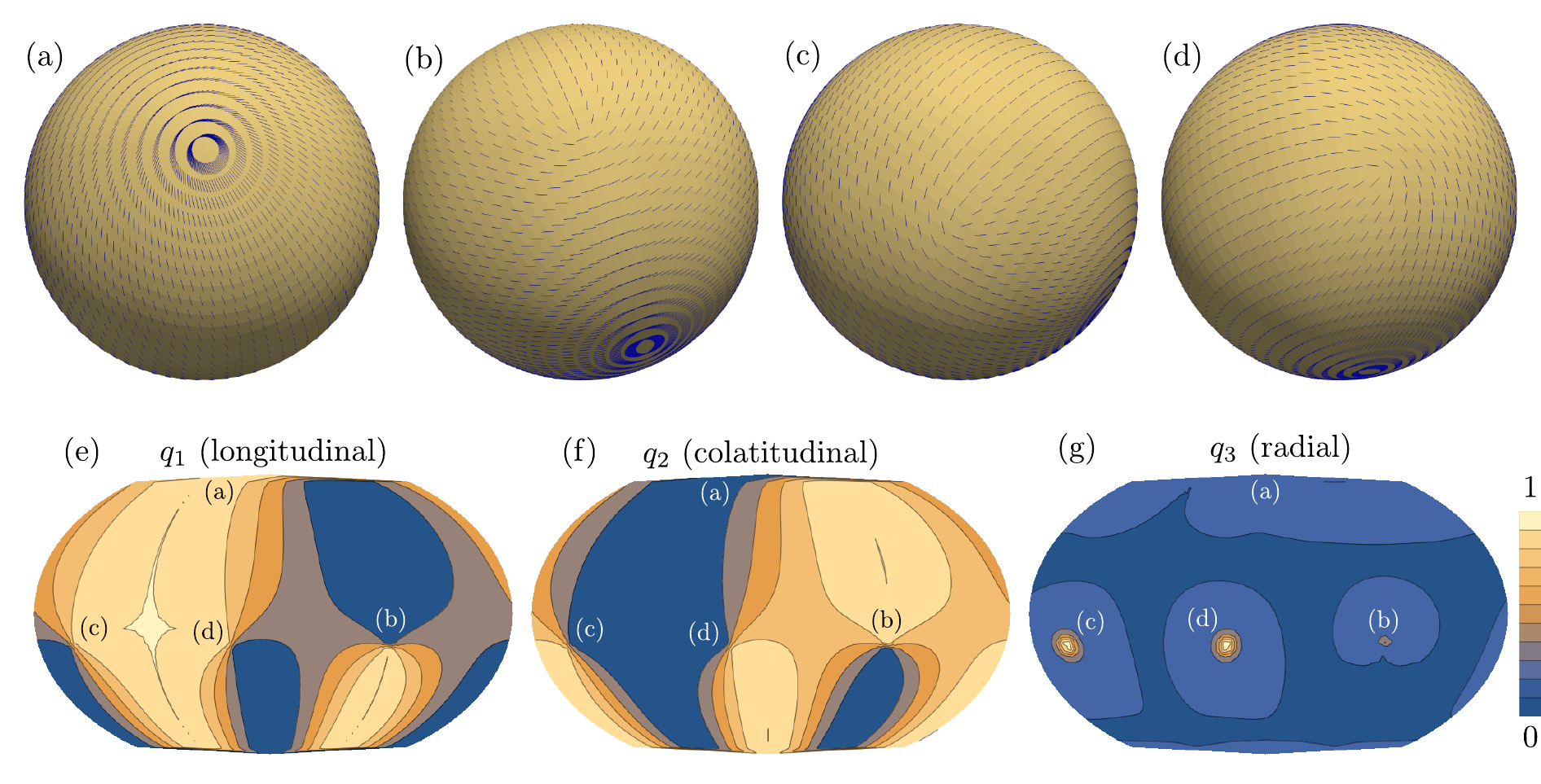}
\caption{\textbf{Tetrahedral configuration  on a sphere of radius $r = 10\lambda$.}  The four  $+1/2$ disclinations are located at vertices of tetrahedron (a--d).  The components  of the alignment tensor $\Nv$ are sketched in (e--g).}
\label{fig:supp-1_2x4}
\end{figure*}

The initial state with four  {$+1/2$ disclinations at the vertices of a tetrahedron}  is given by
\begin{equation}\label{eq:charge_1_2x4-ini}
\sigma_{1/2\times 4} = \dfrac{1}{2} \arg \left( 4ie^{i\phi} \cos^3 \dfrac{\theta}{2}\sin \dfrac{\theta}{2} - \sqrt{2} e^{-2i\phi} \sin^4 \dfrac{\theta}{2} \right).
\end{equation}
The relaxed state is shown in Fig.~\ref{fig:supp-1_2x4}.

\begin{figure*}
\includegraphics[width=\linewidth]{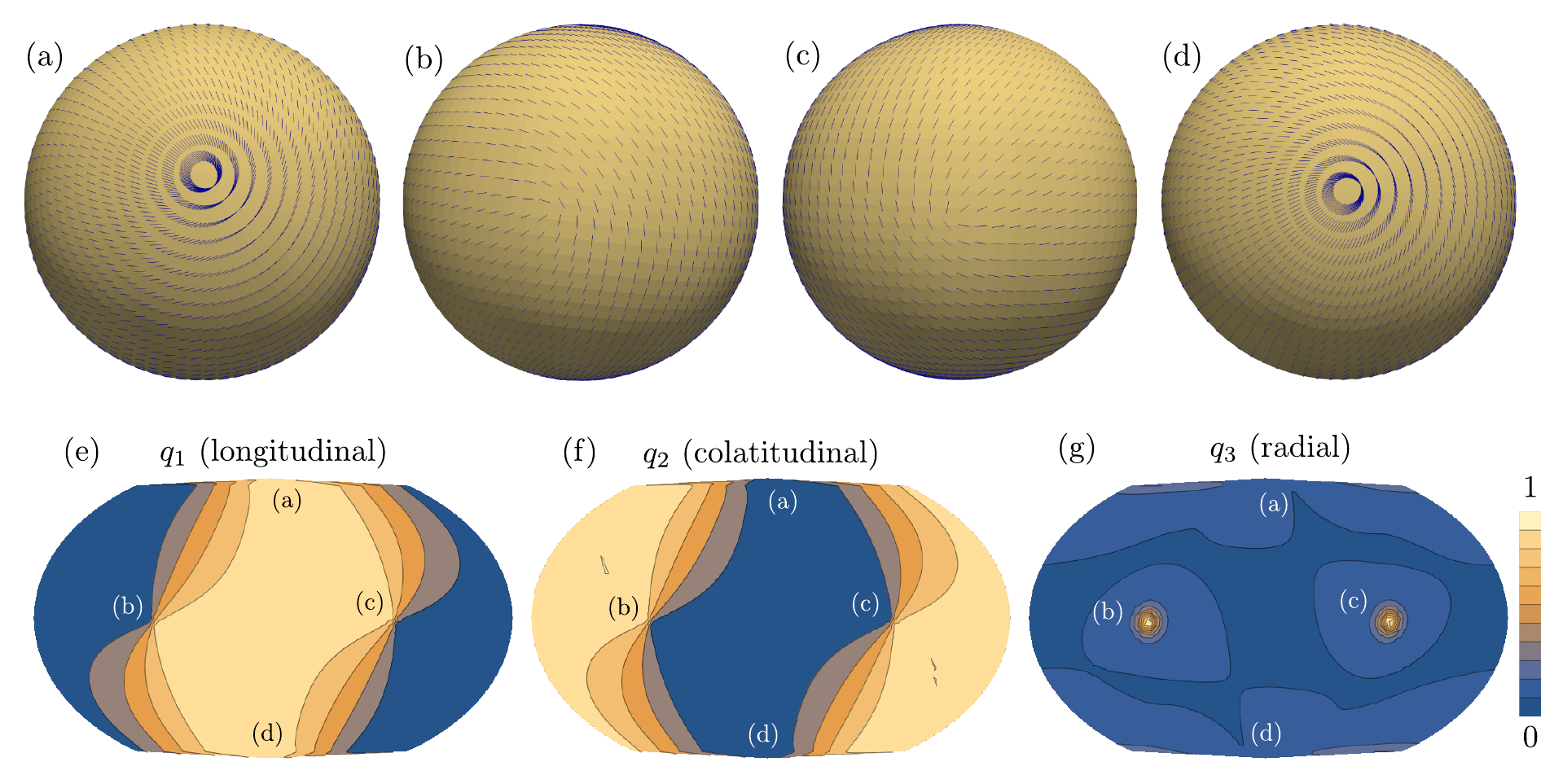}
\caption{  {\textbf{Squared configuration  on a sphere of radius $r = 10\lambda$.}  The four  $+1/2$ disclinations located at the vertices of a square inscribed in a great circle (a--d). The components of the alignment tensor $\Nv$ are sketched in  (e--g). }}
\label{fig:supp-squared}
\end{figure*}

  {The initial state with four $+1/2$ disclinations at the vertices of a square inscribed in  a great circle  is given by}
\begin{equation}\label{eq:charge_squared-ini}
\sigma_{1/2\times 4} = \dfrac{1}{2} \arg \left[ \sqrt{\dfrac{7}{2}} \sin \theta \left( \cos\phi + i \cos\theta\sin\phi \right) \right].
\end{equation}
  {The relaxed state is shown in Fig.~\ref{fig:supp-squared}.}  

The initial state with  {with three disclinations, one with topological charge equal to $+1$ and the other  two  equal to $+1/2$} is in the form
\begin{equation}\label{eq:charge_1_1_2x2-ini}
\sigma_{1+1/2\times 2} = \left[ \dfrac{1}{2} \arg \left( 2ie^{-2i\phi} \sin^4 \dfrac{\theta}{2} - \dfrac{1}{2} \sin^2 \dfrac{\theta}{2} \right) \right].
\end{equation}

\begin{figure*}
\includegraphics[width=0.8\linewidth]{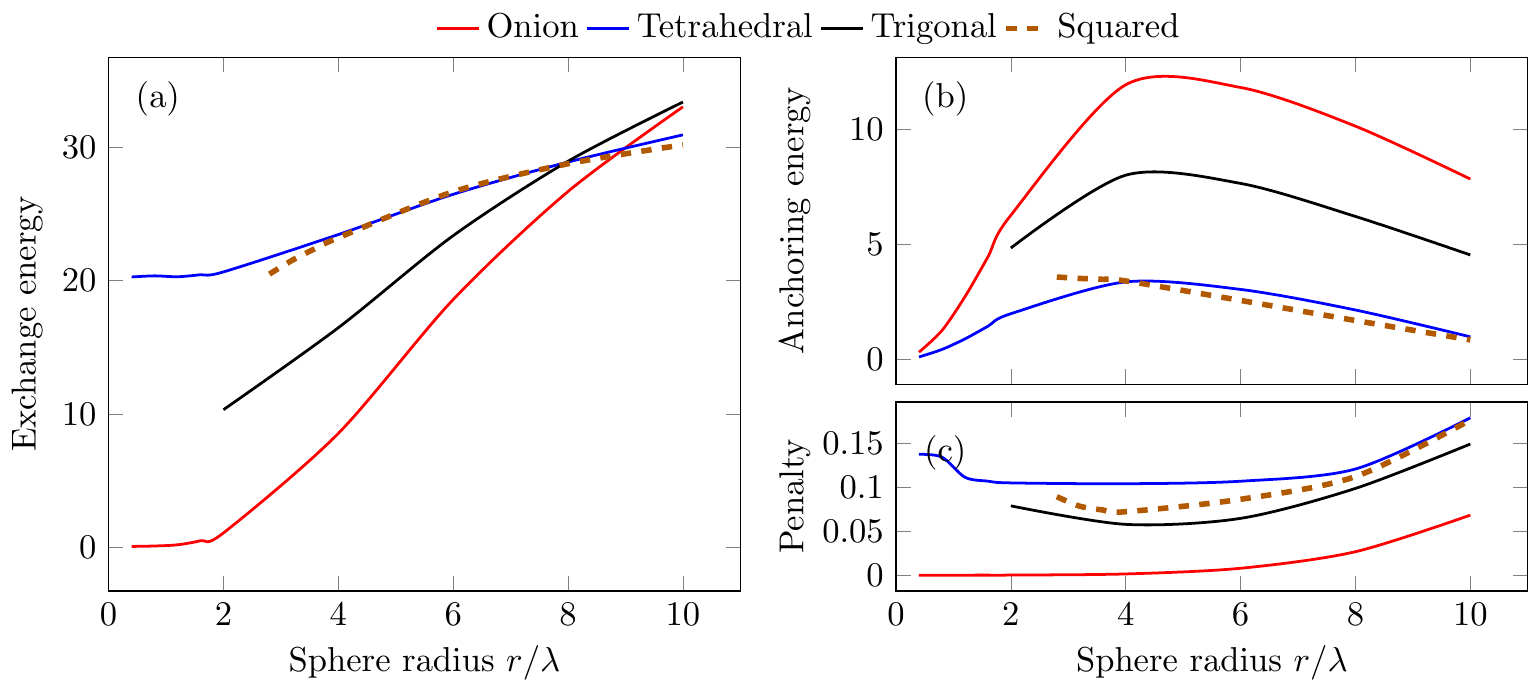}
\caption{  {\textbf{Energy contributions for the different equilibrium configurations.} (a) Exchange energy. (b) Anchoring energy. (c) Penalty $W_p$ to satisfy constraints~\eqref{constr}.}}
\label{fig:supp-energies}
\end{figure*}

 {The exchange and anchoring energies, and penalty $W_p$ of the stable equilibrium configurations are shown in Fig.~\ref{fig:supp-energies}. The stability of the nematic configurations has been  tested with step $\Delta (r/\lambda) = 0.4$ for  $r/\lambda \in [0.4, 4]$, with step $\Delta (r/\lambda) = 2$  for $r/\lambda \in [4, 10]$.} 

 {We conclude by observing that for $r/\lambda \lesssim 2$ the exchange energy of the onion configuration is small and tends to zero as $r/\lambda$ approaches zero.  This means that for small values of $r/\lambda$ the director field tends to be uniform in space.}

\begin{figure*}
\includegraphics[width=\linewidth]{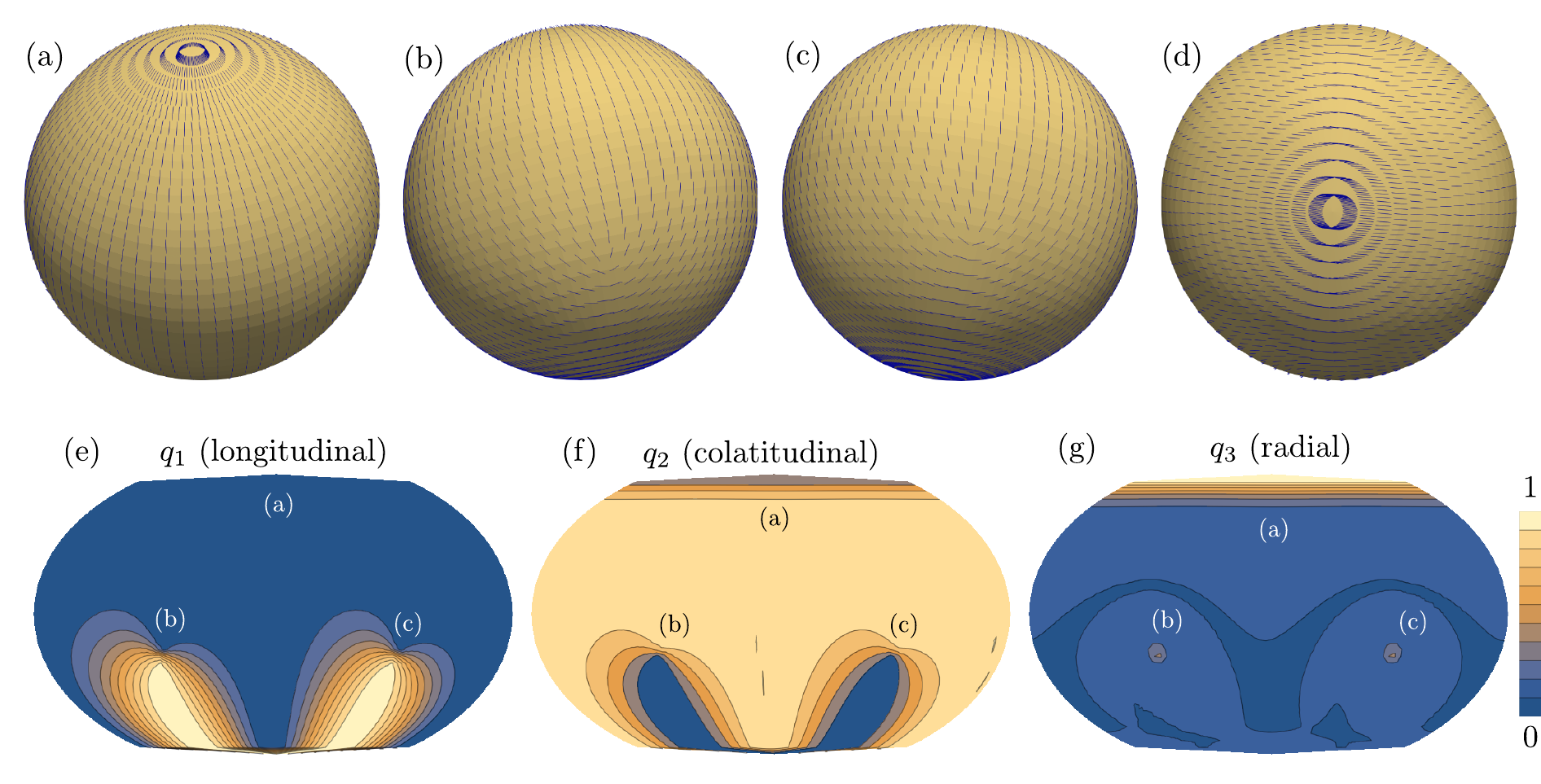}
\caption{\textbf{Trigonal configuration  on a sphere of radius $r = 8\lambda$.}  One $+1$ disclination and  two  $+1/2$ disclinations are located at the vertices of an isosceles triangle inscribed in a great circle (a,b,c). (d) Topologically trivial configuration at the south pole. (e--g) The components of the alignment tensor $\vec{N}$ are sketched in (e,f,g).}
\label{fig:supp-1_1_2x2}
\end{figure*}

